\documentclass[fleqn,twoside,twocolumn,nofootinbib,showkeys]{revtex4} 
\usepackage[sec,nocpr]{ujp} 
\begin{document}
\title[A look at multiplicity distributions via modified combinants]%
{A LOOK AT MULTIPLICITY DISTRIBUTIONS VIA MODIFIED COMBINANTS}%
\author{M.~Rybczy\'nski}
\affiliation{Institute of Physics, Jan Kochanowski University}
\address{25-406 Kielce, Poland}
\email{maciej.rybczynski@ujk.edu.pl and zbigniew.wlodarczyk@ujk.edu.pl}
\author{G.~Wilk}
\affiliation{National Centre for Nuclear Research}
\address{02-093 Warsaw, Poland}
\email{grzegorz.wilk@ncbj.gov.pl}
\author{Z.~W\l odarczyk}
\affiliation{Institute of Physics, Jan Kochanowski University}
\address{25-406 Kielce, Poland}

\udk{539} \pacs{13.85.Hd, 25.75.Gz} \razd{\seci}

\autorcol{M.~Rybczy\'nski, G.~Wilk, Z.~W\l odarczyk}

\setcounter{page}{1}%

\begin{abstract}
The experimentally measured multiplicity distributions exhibit, after closer inspection, peculiarly enhanced void probability and oscillatory behavior of the modified combinants. We show that both these features can be used as additional sources of information, not yet fully explored, on the mechanism of multiparticle production. We provide their theoretical understanding within the class of compound distributions.
\end{abstract}

\keywords{multiplicity distributions, combinants, void probabilities, compound distributions.}

\maketitle

\section{Introduction}

The experimentally measured (non-single diffractive (NSD) charged) multiplicity distributions, $P(N)$ (which are one of the most thoroughly investigated and discussed sources of information on the mechanism of the production process \cite{Kittel}), exhibit, after closer inspection, a peculiarly enhanced void probability, $P(0) > P(1)$ \cite{EVP,Void}, and oscillatory behavior of the so-called modified combinants, $C_j$, introduced by us in \cite{JPG,IJMPA} (and thoroughly discussed in \cite{MRGWZW-EPJC,MRGWZW-PRD}; they are closely connected with the combinants $C_j^{\star}$ introduced in \cite{Combinants-1} and  discussed occasionally for some time \cite{CombUse1,CombUse1a,CombUse1b,CombUse2b,AIP-B,CombUse5}). Both features were only rarely used as a source of information. We demonstrate that the modified combinants can be extracted experimentally from the measured $P(N)$ by means of a recurrence relation involving all $P(N<j)$, and that new information is hidden in their specific distinct oscillatory behavior, which, in most cases, is not observed in the $C_j$ obtained from the $P(N)$ commonly used to fit experimental results \cite{JPG,IJMPA,MRGWZW-EPJC,MRGWZW-PRD}. We discuss the possible sources of such behavior and the connection of the $C_j$ with the enhancement of void probabilities, and their impact on our understanding of the multiparticle production mechanism, with emphasis on understanding both phenomena within the class of compound distributions.

\section{Recurence relation and modified combinants}

The dynamics of the multiparticle production process is hidden in the way in which the consecutive measured multiplicities $N$ are connected. There are two ways of characterizing multiplicity distributions: by means of generating functions, $G(z) = \sum_{N=0}^{\infty} P(N) z^N$, or by some form of recurrence relation between the $P(N)$'s. In the first case, one uses as a reference the Poisson distribution and characterizes departures from it by means of combinants $C_N^{\star}$ defined as \cite{Combinants-1}
\begin{equation}
C^{\star}_j = \frac{1}{j!} \frac{d^j \ln G(z)}{d z^j}\bigg|_{z=0}, \label{C_j_star}
\end{equation}
or by the expansion
\begin{equation}
\ln G(z) = \ln P(0) + \sum^{\infty}_{j=1} C^{\star}_j z^j. \label{CombDef}
\end{equation}
For a Poisson distribution $C_1^* =\langle N\rangle$ and $C^{*}_{j>1}=0$. The combinants were used in the analysis of experimental data in \cite{CombUse1,CombUse1a,CombUse1b,CombUse2b,AIP-B,CombUse5}. In \cite{CombUse1a,AIP-B} it was demonstrated that they are particularly useful in identifying the nature of the emitting source. It turns out that in the case of $S$ sources emitting particles without any restrictions concerning their number, the multiplicity $P^S(N)$ is a completely symmetric function of degree $N$ of the probabilities of emission, $p_i$,  the generating function of which reduces for $p_i \to 0$ to the generating function of the Poisson Distribution (PD) and, for all probabilities remaining the same, $p_i = p$, it reduces to the generating function of the Negative Binomial Distribution (NBD). In this case the combinants are given by a power series
\begin{equation}
C_j^{\star} = \frac{1}{j}\sum_{i=1}^S p_i^j \label{NBDC}
\end{equation}
and are always positive. However, when  each of the sources can emit only a given number of particles (let us assume, for definiteness, that at most only one particle), then $P^S(N)$ is an elementary symmetric function of degree $N$ in the arguments and the corresponding combinants are given by
\begin{equation}
C_j^{\star} = (-1)^{j+1} \frac{1}{j} \sum_{i=1}^S \left( \frac{p_i}{1-p_i}\right )^j, \label{BDC}
\end{equation}
and alternate in sign for different $j$'s. For all probabilities remaining the same, $p_i = p$, a generating function in this scenario reduces to the generating function of the Binomial Distribution (BD) and the combinants oscillate rapidly with period equal to $2$.

Note that in both cases we were working with probabilities $p_i$ which were not extracted from experiment but whose values were taken in such a way as to reproduce the measured multiplicity distributions. They are then usually represented by one of the known theoretical formulae for multiplicity distributions, $P(N)$, which can be defined either by the generating functions mentioned above or by some recurrence equations connecting different $P(N)$. In the simplest (and most popular) case one assumes that the multiplicity $N$ is directly influenced only by its neighboring multiplicities, $(N \pm 1)$, i.e., that:
\begin{equation}
(N+1)P(N+1) = g(N)P(N),\quad g(N) = \alpha + \beta N. \label{rr1}
\end{equation}
From this recurrence equation emerge the BD (when $\alpha = Kp/(1-p)$ and $\beta = -\alpha/K$), the PD (when $\alpha = \lambda$ and $\beta = 0$), and the NBD (when $\alpha = kp$ and $\beta = \alpha/k$, where $p$ denotes the probability of particle emission). Usually the first choice of $P(N)$ in fitting data is a single NBD \cite{KNO}, or two \cite{GU,PG}, three \cite{Z}, or multi-component NBDs \cite{DN} (or some other forms of $P(N)$ \cite{Kittel,KNO,MF}). However, such a procedure only improves the agreement at large $N$, whereas the ratio $R = data/fit$ still deviates dramatically from unity at small $N$ for all fits \cite{JPG,IJMPA}. This means that the measured $P(N)$ contains information which is not yet captured by the rather restrictive recurrence relation (\ref{rr1}). Therefore, in \cite{JPG} we proposed to use a more general form of the recurrence relation (used, for example, in counting statistics when dealing with multiplication effects in point processes \cite{ST}):
\begin{equation}
(N + 1)P(N + 1) = \langle N\rangle \sum^{N}_{j=0} C_j P(N - j). \label{rr2}
\end{equation}
This relation connects multiplicities $N$ by means of some coefficients $C_j$, which contain the memory of particle $N+1$ about all the $N-j$ previously produced particles. The most important feature of this recurrence relation is that $C_j$ can be directly calculated from the experimentally measured $P(N)$ by reversing Eq. (\ref{rr2}) \cite{JPG,IJMPA,MRGWZW-EPJC,MRGWZW-PRD}:
\begin{equation}
\langle N\rangle C_j = (j+1)\left[ \frac{P(j+1)}{P(0)} \right] - \langle N\rangle \sum^{j-1}_{i=0}C_i \left[ \frac{P(j-i)}{P(0)} \right]. \label{rCj}
\end{equation}

The modified combinants $C_j$ defined by the recurrence relation (\ref{rCj}) are closely related to the combinants $C^{\star}_j$ defined by Eq. (\ref{C_j_star}), namely
\begin{equation}
C_j = \frac{j+1}{\langle N\rangle} C^{\star}_{j+1}. \label{connection}
\end{equation}
Using Leibnitz's formula for the $j^{th}$ derivative of the quotient of two functions $x=G'(z)/G(z)$,
\begin{equation}
x^{(j)}= \frac{1}{G}
\left(G'^{(j)}-j!\sum_{k=1}^{j} \frac{G'^{(j+1-k)}}{(j+1-k)!}\frac{x^{(k-1)}}{(k-1)!} \right),
\label{derivate}
\end{equation}
where $G'(z)/G(z)=[lnG(z))]$ and $G(z)^{(N)}/N!\vert_{z=0}=P(N)$, we immediately obtain the recurrence relation (\ref{rCj}).

The modified combinants, $C_j$, share with the combinants $C_j^{\star}$ the apparent ability of identifying the nature of the emitting source mentioned above (with, respectively, Eq. (\ref{NBDC}) corresponding to the NBD case with no oscillations, and Eq. (\ref{BDC}) corresponding to the rapidly oscillating case of a BD). This also means that $C_j$ can be calculated from the generating function $G(z)$ of $P(N)$,
\begin{equation}
\langle N\rangle C_j = \frac{1}{j!} \frac{ d^{j+1} \ln G(z)}{d z^{j+1}}\bigg|_{z=0}. \label{GF_Cj}
\end{equation}
Thus, whereas the recurrence relation, Eq. (\ref{rCj}), allows us to obtain the $C_j$ from the experimental data on $P(N)$, Eq. (\ref{GF_Cj}) allows for their calculation from the distribution defined by the generating function $G(z)$.

Note that the $C_j$ provide a similar measure of fluctuations as the set of cumulant factorial moments, $K_q$, which are very sensitive to the details of the multiplicity distribution and are frequently used in phenomenological analyses of data (cf., \cite{Kittel,Book-BP}),
\begin{equation}
K_q = F_q - \sum_{i=1}^{q-1}\binom{q-1}{i-1} K_{q-i}F_i, \label{cumfactmom}
\end{equation}
where
$F_q = < N(N-1)(N-2)\dots(N-q+1)>$
are the factorial moments. The $K_q$ can be expressed as an infinite series of the $C_j$,
\begin{equation}
K_q = \sum_{j=q}^{\infty}\frac{(j-1)!}{(j-q)!}\langle N\rangle C_{j-1}. \label{KconC}
\end{equation}
However, while the cumulants are best suited to study densely populated regions of phase space,  combinants are better suited for the study of sparsely populated regions because, according to Eq. (\ref{rCj}), calculation of $C_j$ requires only a finite number of probabilities $P(N<j)$ (which may be advantageous in applications).

\begin{figure}[h]
\begin{center}
\includegraphics[scale=0.7]{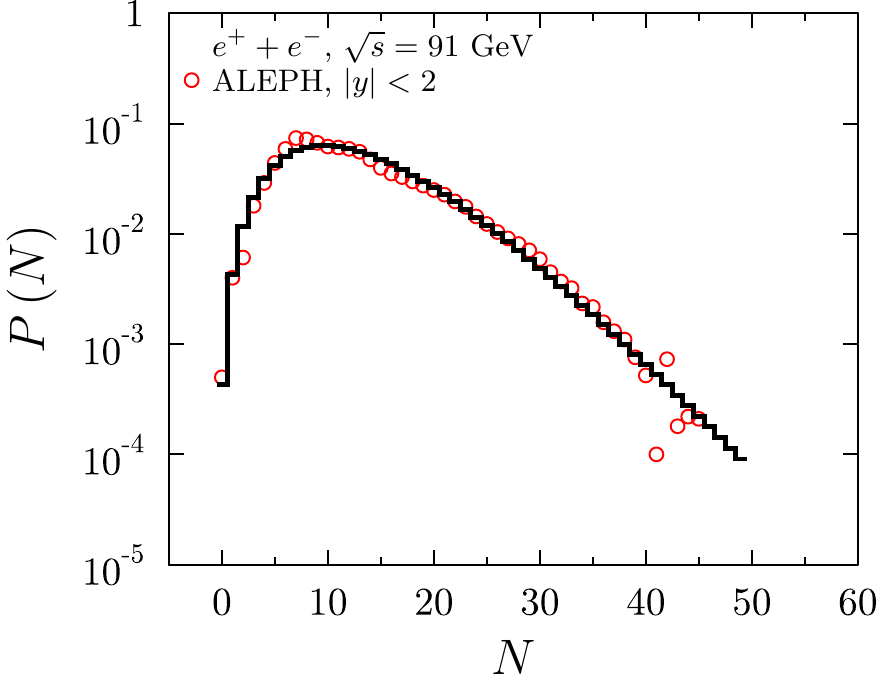}\\
\includegraphics[scale=0.7]{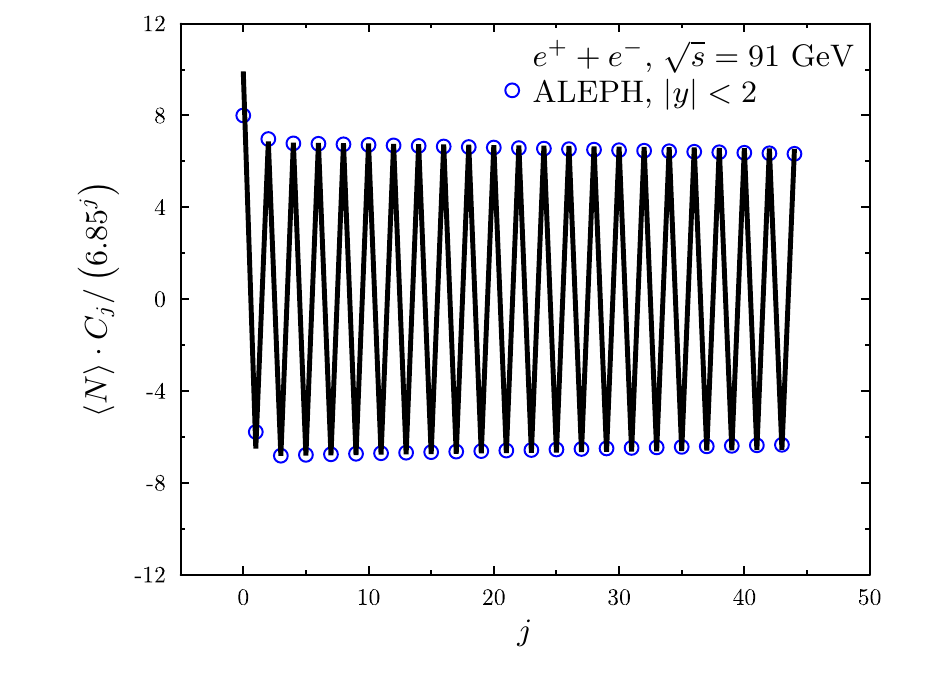}
\end{center}
\vspace{-5mm}
\caption{Upper panel: Data on $P(N)$ measured in $e^+e^-$ collisions by the ALEPH experiment at $91$ GeV \cite{ALEPH} are fitted by the distribution obtained from the generating function given by the product $G(z)=G_{BD}(z)G_{NBD}(z)$
with parameters: $k'=1$ and $p'=0.8725$ for the BD and $k=4.2$ and $p=0.75$ for the NBD.
Lower panel: the modified combinants $C_j$ deduced from these data on $P(N)$. They can be fitted by $C_j$ obtained from the same generating function  with the same parameters as used for fitting $P(N)$. }
\label{F1}
\end{figure}

The modified combinants share with the cumulants the property of additivity. For a random variable composed of independent random variables, with its generating function given by the product of their generating functions, $G(x)=\prod_jG_j(x)$, the corresponding modified combinants are given by the sum of the independent components. To illustrate this property let us consider the $e^+e^-$ data and use the generating function $G(z)$ formally treated as a generating function of the multiplicity distribution $P(N)$ in which $N$ consists of both the particles from the BD ($N_{BD}$) and from the NBD ($N_{NBD}$):
\begin{equation}
N = N_{BD} + N_{NBD}. \label{NN}
\end{equation}
In this case the multiplicity distribution can be written as
\begin{equation}
P(N) = \sum_{i=0}^{min\left\{ N,k'\right\}} P_{BD}(i)P_{NBD}(N-i) \label{PbdPnbd}
\end{equation}
and the respective modified combinants  as
\begin{equation}
\langle N\rangle C_j = \left< N_{BD}\right>C_j^{(BD)} + \left< N_{NBD}\right> C_j^{(NBD)}. \label{Cjbdnbd}
\end{equation}
Fig. \ref{F1} shows the results of attempts to fit both the experimentally measured \cite{ALEPH} multiplicity distributions and the corresponding modified combinants $C_j$ calculated from these data (cf. \cite{EPEM} for details).
The fits shown in Fig. \ref{F1} correspond to parameters: $k'=1$ and $p'=0.8725$ for the BD and $k=4.2$ and $p=0.75$ for the NBD.

Concerning the void probabilities, at all energies of interest one observes that $P(0) > P(1)$, a feature which cannot be reproduced by any composition of the NBD used to fit the data \cite{MRGWZW-PRD}. To visualize the importance of this result note first that $P(0)$ is strongly connected with the modified combinants $C_j$, in fact
\begin{equation}
P(0) = \exp\left( - \sum_{j=0}^{\infty} \frac{\langle N\rangle}{j+1}C_j\right). \label{P(0)}
\end{equation}
From Eq. (\ref{rCj}) one can deduce that the $P(0) > P(1)$ property is possible only when $\langle N\rangle C_0 < 1$.
For most multiplicity distributions also $P(2) > P(1)$, which results in an additional condition, $C_1 > C_0 \left( 2 - \langle N\rangle C_0\right)$; taken together this means that in this case $C_1 > C_0$.
However, because of the normalization condition, $\sum_{j=0}^{\infty} C_j =1$, such an initial increase of $C_j$  cannot continue for all ranks $j$ and we should observe some kind of nonmonotonic behavior of $C_j$ with rank $j$ in this case. This means that  all multiplicity distributions for which the modified combinants $C_j$ decrease monotonically with rank $j$ do not exhibit the enhanced void probability.

\section{Compound distributions}

\begin{figure}[h]
\begin{center}
\includegraphics[scale=0.38]{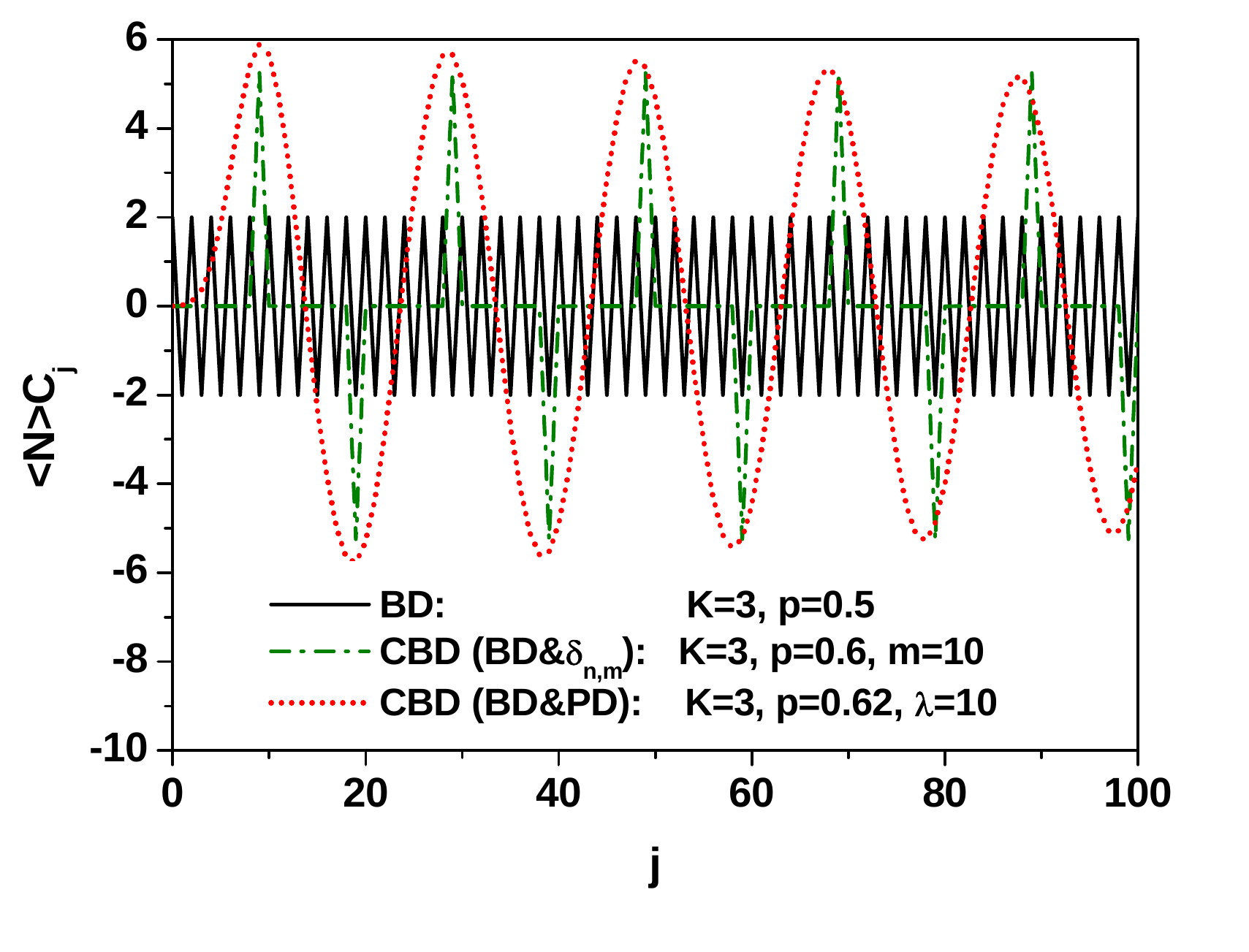}
\end{center}
\vspace{-10mm}
\caption{ $C_j$ for BD, BD compounded with $\delta_{n,m}$ with $m=10$ and compounded with a Poison distribution with $\lambda =10$.}
\label{F2}
\end{figure}

To continue we use the idea of {\it compound distributions} (CD) which are applicable when (as in our case) the production process consists of a number $M$ of some objects (clusters/fireballs/etc.) produced according to a distribution $f(M)$ (defined by a generating function $F(z)$), which subsequently decay independently into a number of secondaries, $n_{i = 1,\dots, M}$, following some other (always the same for all $M$) distribution, $g(n)$ (defined by a generating function $G(z)$). The resultant multiplicity distribution,
\begin{equation}
h\left( N =\sum_{i=0}^M n_i\right) = f(M)\otimes  g(n), \label{ftimesg}
\end{equation}
is a compound distribution of $f$ and $g$ with generating function
\begin{equation}
H(z) = F[G(z)]. \label{CD_GF}
\end{equation}
Eq. (\ref{CD_GF}) means that in the case where $f(M)$ is a Poisson distribution with generating function
\begin{equation}
F(z) = \exp[\lambda (z - 1)], \label{GFPD}
\end{equation}
then, for any other distribution $g(n)$ with generating function $G(z)$, the combinants obtained from the compound distribution $h(N) = P_{PD} \otimes g(n)$ and calculated using Eq. (\ref{GF_Cj}), do not oscillate and are equal to
\begin{equation}
C_j = \frac{\lambda (j+1)}{\langle N\rangle}g(j+1). \label{Cjcalculated}
\end{equation}
This fact explains why $C_j$ from NBDs do not oscillate. This is because the NBD is the compound distribution of poisson and logarithmic distributions. This means that $g(n) = - p^n/[n\ln(1-p)]$ and $h(N)$ is the NBD with $k= - \lambda /\ln(1-p)$. In this case the $C_j$ coincide with those derived before and given by Eq. (\ref{NBDC}). Actually, this reasoning applies to all more complicated compound distributions, with any distribution itself being a compound poisson distribution. This property limits the set of distributions $P(N)$ leading to oscillating $C_j$ to a BD and to all compound distributions based on it. In this case the period of the oscillations is determined by the number of particles emitted from the source. Whereas for compound distributions based on the BD with  $P(n)=\delta_{n,m}$ we have
\begin{equation}
C_j = (-1)^{j/m+1} \frac{K}{\langle N\rangle} \left( \frac{p}{1 - p}\right)^{j/m+1} , \label{CjBD}
\end{equation}
(for $j=mk$ and $C_j=0$ for $j\neq mk$, where $k=1,2,3,...$), for broader distributions $P(n)$ we get a smoother $C_j$ dependence on rank $j$. For example, for $P(n)$ given by a Poissonian distribution (with expected value $\lambda$) we obtain a Compound Binomial Distribution (CBD) with generating function
\begin{equation}
H(z) = \left\{ p \exp[ \lambda (z-1)] + 1 -p \right\}^K. \label{H_FG}
\end{equation}
and the modified combinants are given by
\begin{equation}
C_j =  \frac {(-1)^{j+1}K e^{\lambda} \lambda^{j+1}\frac {1-p}{p}}{\langle N\rangle \left(e^{\lambda}
\frac{1-p}{p}+1 \right)^{j+1}}
A_j \left( e^\lambda \frac{p-1}{p} \right )
\label{CjBDPoisson}
\end{equation}
where $A_j(x)$ are the Eulerian polynomials. As an illustration we show in Fig. \ref{F2} that compounding a BD with a Poisson distribution one gains control of both the period of the oscillations (now equal to $2\lambda$) and their amplitude. However, it turns out that such a combination does not allow us to fit data.

\section{Multi - component}

\begin{figure}[h]
\begin{center}
\includegraphics[scale=0.7]{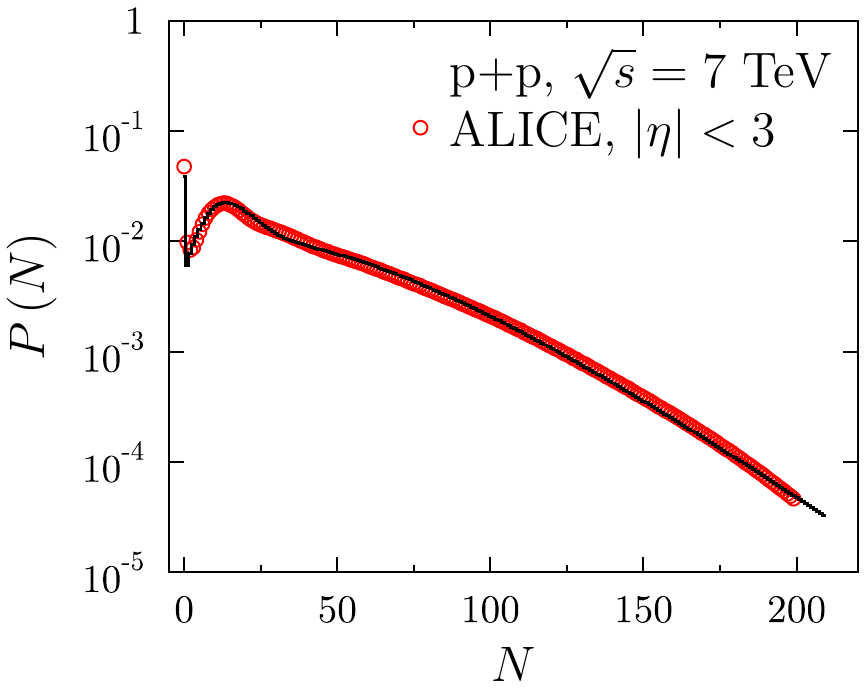}\\
\includegraphics[scale=0.7]{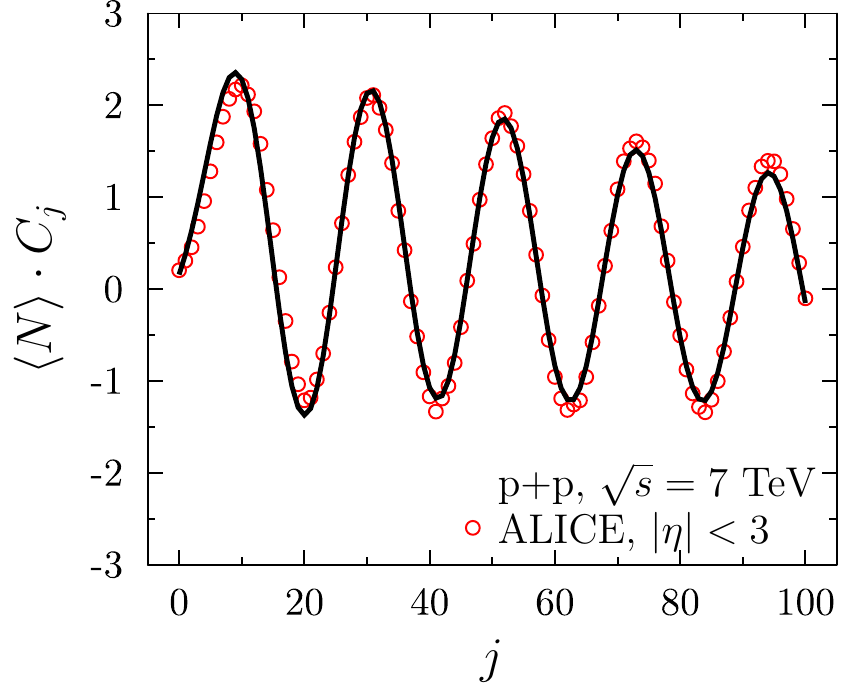}
\end{center}
\vspace{-5mm}
\caption{ Multiplicity distributions $P(N)$ measured in $pp$ collisions by ALICE \cite{ALICE} (upper panel) and  the corresponding modified combinants $C_j$ (lower panel). Data are fitted using a two compound distribution (BD+NBD) given by Eqs. (\ref{2-CBD}) and (\ref{2CBD}) with parameters:
 $K_1 = K_2 = 3$, $p_1 = 0.9$, $p_2 = 0.645$, $k_1 = 2.8$, $k_2 = 1.34$, $m_1 = 5.75$, $m_2 = 23.5$ , $w_1 = 0.24 $ and $w_2 = 0.76$ .}
\label{F3}
\end{figure}

The situation improves substantially when one uses a multi-CBD based on Eq. (\ref{H_FG}), but still the agreement is not satisfactory. It turns out that the situation improves dramatically if one replaces the Poisson distribution by a NBD and, additionally, uses a two-component version of such a CBD, with
\begin{equation}
P(N) = \sum_{i=1,2} w_i h\left(N; p_i, K_i, k_i, m_i\right)
\label{2CBD}
\end{equation}
with the generating function of each component equal to
\begin{equation}
H(z) = \left[ p\left( \frac{1 - p'}{1 - p'z}\right)^k + 1 - p\right]^K. \label{2-CBD}
\end{equation}
In such a case, as can be seen in Fig. \ref{F3}, one gains satisfactory control over both the periods of the oscillations and their amplitudes, and on their behavior as a function of the rank $j$, and one can nicely fit both the $P(N)$ and $C_j$.  Of special importance is the fact that the enhancement $P(0)>P(1)$ is also reproduced in this approach.

The above result also explains the apparent success in fitting the experimentally observed oscillations of $C_j$ by using a weighted sum of the three NBD used in \cite{Zborovsky}. Such a distribution uses freely selected weights and parameters $(p,k)$ of the NBDs and therefore resembles the compound distribution of the BD with the NBD. However, note that the sum of $M$ variables (with $M = 0,~1,~2,\dots$), each from the NBD characterized by parameters $(p,k)$, is described by a NBD characterized by $(p,Mk)$, therefore, as discussed before, it cannot reproduce the void probability $P(0)$. This can be reproduced only in the case where $M = 0,~1,\dots, K$ is distributed according to a BD and we have a $K$-component NBD (where the consecutive NBDs have precisely defined parameters $k$),
\begin{equation}
P(N) = \sum_{M=0}^K P_{BD}(M) P_{NBD}(N;p,Mk), \label{BD-NBD}
\end{equation}
This is because, in this case, one also has the $M=0$ component, which is lacking in the previous multi-NBD case used in \cite{Zborovsky}. This is the reason that, whereas the compound $(BD\&NBD)$ distribution reproduces the void probability, $P(0)$, the single NBD (or any combination of NBDs) do not. This means that the observation of the peculiar behavior of the void probability discussed above signals the necessity of using some compound distribution based on the BD to fit data for $P(N)$ (and the $C_j$ obtained from it).

\section{Summary and conclusions}

Since the time of Ref. \cite {Combinants-1} one encounters essentially no detailed experimental studies of the combinants and only rather sporadic attempts at their phenomenological use to describe the multiparticle production processes.
We demonstrate that the modified combinants $C_j$ are a valuable tool for investigations of multiplicity distributions, and  $C_j$ deduced from the measured multiplicity distributions, P(N), could provide additional information on the dynamics of the particle production. This, in turn, could allow us to reduce the number of possible interpretations presented so far and, perhaps, answer some of the many still open fundamental questions (that this is possible despite experimental errors has been shown in \cite{Zborovsky,MRGWZW-PRD}). Finally, let us note that a large number of papers suggest some kind of universality in the mechanisms of hadron production in $e^+e^-$ anihilations and in $pp$ and $p\bar{p}$ collisions. This arises from observations of the average multiplicities and relative dispersions in both types of processes (cf. for example, \cite{Biswas,G-OR}). However, as we have shown here, the modified combinant analysis reveals differences between these processes. Namely, while in $e^+e^-$ anihilations  we observe oscillations of $C_j$ with period $2$, in  $pp$  collisions the period of oscillation is $\sim 10$ times longer and the amplitude of the oscillations in both types of processes differs dramatically. At the moment this problem remains open and awaits further investigation.

\vskip3mm \textit{This research  was supported in part by the Polish Ministry of Science and Higher Education
(Contract  No. DIR/WK/2016/2010/17-1) and the National Science Centre (NCN) (Contract No. DEC-2016/22/M/ST/00176) (G. W.) and
by the NCN grant 2016/23/B/ST2/00692 (M. R.). We would like to thank Dr Nicholas Keeley for reading the manuscript.}

\end{document}